\documentclass[showkeys,showpacs,superscriptaddress]{revtex4}
\usepackage{amsmath}
\usepackage{amssymb}
\usepackage{graphicx}
\usepackage{color,ulem}
\allowdisplaybreaks

\begin{document}

\title{
Thick branes in higher-dimensional $f(R)$ gravity
}

\author{
Vladimir Dzhunushaliev
}
\email{v.dzhunushaliev@gmail.com}
\affiliation{
Department of Theoretical and Nuclear Physics,  Al-Farabi Kazakh National University, Almaty 050040, Kazakhstan
}
\affiliation{
Institute of Experimental and Theoretical Physics,  Al-Farabi Kazakh National University, Almaty 050040, Kazakhstan
}
\affiliation{
Academician J.~Jeenbaev Institute of Physics of the NAS of the Kyrgyz Republic, 265 a, Chui Street, Bishkek 720071, Kyrgyzstan
}
\affiliation{
Institut f\"ur Physik, Universit\"at Oldenburg, Postfach 2503
D-26111 Oldenburg, Germany
}

\author{Vladimir Folomeev}
\email{vfolomeev@mail.ru}
\affiliation{
Institute of Experimental and Theoretical Physics,  Al-Farabi Kazakh National University, Almaty 050040, Kazakhstan
}
\affiliation{
Academician J.~Jeenbaev Institute of Physics of the NAS of the Kyrgyz Republic, 265 a, Chui Street, Bishkek 720071, Kyrgyzstan
}
\affiliation{
Institut f\"ur Physik, Universit\"at Oldenburg, Postfach 2503
D-26111 Oldenburg, Germany
}

\author{Galia Nurtaeva}
\affiliation{
Department of Theoretical and Nuclear Physics,  Al-Farabi Kazakh National University, Almaty 050040, Kazakhstan
}

\author{Sergei D. Odintsov}
\affiliation{Institut de Ciencies de lEspai (IEEC-CSIC),
Carrer de Can Magrans, s/n, 08193 Barcelona, Spain
}
\affiliation{Instituci\'{o} Catalana de Recerca i Estudis Avan\c{c}ats (ICREA),
Passeig Llu\'{\i} Companys, 23 08010 Barcelona, Spain
}


\begin{abstract}
We study the thick brane model within $f(R)\sim R^n$ modified gravity in $D$-dimensional spacetimes with $D\geq 6$.
The system under consideration consists of two branes orthogonal to each other:
the four-dimensional Lorentzian brane and $(D-5)$-dimensional Euclidean one.
It is numerically shown that, for a given $D$, regular vacuum asymptotically anti-de Sitter solutions exist only in the range $1<n<D/2$.
Depending on the values of $n$ and boundary conditions imposed on the Lorentzian brane, the solutions can pass or not pass through a fixed point
located on the Lorentzian brane, and also be $Z_2$-symmetric or nonsymmetric.
In the large-$D$ limit, we find approximate analytic solutions.
It is also shown that a test scalar field is trapped on
the Lorentzian brane at any $D$.
\end{abstract}

\pacs{04.50.--h
}

\keywords{
Higher-dimensional thick branes, modified gravity
}

\maketitle

\section{Introduction}
Almost immediately upon creation of Einstein's general relative (GR) the question about the dimensionality of the world in which we live appeared.
While within GR the world is usually assumed to be four-dimensional, it is not impossible that extra dimensions, by some reason hidden
from direct observation, may also exist. Such dimensions can manifest themselves in a variety of ways,
as it takes place, for example, in the five-dimensional Kaluza-Klein theory that is equivalent to four-dimensional Einstein gravity
coupled to Maxwell's electromagnetism~\cite{wesson}. Such a possibility of the unification of fundamental interactions within
multidimensional theories  has been further developed in superstring theories, where it is assumed that all four known
interactions  of our four-dimensional spacetime result from the spontaneous compactification of a higher-dimensional space~\cite{wesson}.
	
An alternative approach to explain the unobservability of extra dimensions, called brane world scenario, assumes the existence of noncompact extra dimensions.
Within this approach, it is supposed that particles corresponding to electroweak and strong interactions are confined on some hypersurface
(called a brane) which in turn is embedded in a multidimensional space (called a bulk). In this case only gravitational field and some exotic forms of matter (e.g., a dilaton field)
can propagate in the bulk. It is assumed that our Universe is such a brane-like object~\cite{akama,rubakov}.
The use of the brane world scenario permits one to solve, in particular,
the known hierarchy problem in high-energy physics~\cite{arkani1,arkani2,gog,rs} and to investigate other open questions in particle physics and cosmology,
such as the fermion generation puzzle \cite{fermi1,fermi11,fermi2} or the nature of dark energy \cite{de} and dark matter \cite{dm}.

The literature in the field offers two types of branes -- thin and thick ones. They differ in principle by the behavior of the warp factor  $a(z)$ [see Eq.~\eqref{metr_gen} below]:
in the thin-brane case, a discontinuity of the derivative  $da/dz$ occurs, and in the thick-brane case the solution is smooth.
For five-dimensional problems, the presence of such a discontinuity is usually not a problem,
as long as one is interested in the scales larger than the brane thickness. But in higher-dimensional cases the self-interaction of bulk matter (e.g., gravitation)
can lead to (unavoidable) divergences that does not allow to put any kind of matter on a brane~\cite{Dzhunushaliev:2009va}.
For this reason, in considering  higher-dimensional problems, one has to refuse the thin brane approximation in favour of more realistic models with finite
brane thickness.

Thick brane solutions can be obtained both within GR~\cite{Dzhunushaliev:2009va} and in modified gravity theories~\cite{Liu:2017gcn}.
Various types of modified gravities are used quite widely both in modeling the early stages of the evolution of the Universe and in explaining its current accelerated expansion
(for a general review on the subject, see, e.g., Refs.~\cite{Capozziello:2010zz,Nojiri:2010wj,delaCruzDombriz:2012xy,Nojiri:2017ncd}).
In doing so, while to obtain brane solutions in GR  one has to involve some matter
(there are no vacuum brane solutions), in modified gravities it is already not necessary. For example, within $f(R)$ gravities, it was shown
in Refs.~\cite{Dzhunushaliev:2009dt,Zhong:2015pta} that such theories allow the existence of vacuum five-dimensional thick brane solutions.
It appears to be interesting by itself, since regular vacuum solutions can be obtained only in rare cases.
For example, in Maxwell's electrodynamics, there are no regular vacuum solutions (except electromagnetic waves, which, on the other hand,
have infinite energy).
The same situation takes place in GR as well: except gravitational waves,  regular vacuum solutions are absent. Physically it can be easily explained:
in field theories, in order for solutions to exist, an external source is needed; without such a source, regular solutions can apparently be obtained only through the use of nontrivial structure of a field
(solutions like monopoles, instantons, etc.). From the mathematical point of view, this means that in order to obtain regular solutions differential field equations must either have
a source (in topologically trivial cases) or one has to choose special boundary conditions (in topologically nontrivial cases).

In the present paper we demonstrate that within $f(R)$ gravity there are topologically trivial regular solutions that can describe higher-dimensional thick branes.
From the mathematical point of view, their existence is a consequence of the presence of nonlinear higher derivative terms.
From the physical point of view, this means that the corresponding field theories are much more complicated than those containing second derivatives only.
Apparently, here the difference between such theories like Maxwell's electrodynamics, Yang-Mills theory, and GR on the one hand and
modified theories of gravity on the other is manifested.

\section{Equations and solutions}
\label{eqs}
\subsection{General equations}

In general, the action for modified gravity in a $D=(4+m)$-dimensional spacetime can be written in the form [the metric signature is $(+,-,-,\ldots)$]
\begin{equation}
	S = \int d^D x\sqrt {^{(D)} G} \,F(R)
	+ S_{\text{GHY}} + S_1 .
\label{2_10}
\end{equation}
Here $F(R) = - \frac{M^{m+2}}{2}R + f(R)$, where
$f(R)$ is an arbitrary nonlinear function of the scalar curvature $R$ and $M$ is the fundamental mass scale;
$G$ is the determinant of the $D$-dimensional metric $G_{AB}$ (hereafter capital Latin indices refer to a multidimensional spacetime).
The above action contains  the Gibbons-Hawking-York boundary term $S_{\text{GHY}}$,
which in $F(R)$ gravity is \cite{Guarnizo:2010xr}
\begin{equation}
	S_{\text{GHY}} \sim \oint d^{D-1}x  \sqrt{|h|}F_R K,
\label{GHY_term}
\end{equation}
where $h$ is the determinant of the induced metric
$h_{\mu \nu} = G_{\mu \nu} - n_\mu n_\nu$
on the boundary; $n_\mu$ is a unit vector normal to the brane and directed into the bulk;
$K$ is the trace of the extrinsic curvature of the boundary.
The term  $S_{\text{GHY}}$ is needed to eliminate surface terms that appear when one varies the action in deriving field equations.
It is also important in quantum gravity.

The action~\eqref{2_10} also contains the surface term $S_1$, which can be employed in constructing thin brane models where
matter is confined on an infinitely thin (domain) wall (see Appendix~\ref{sur_act}).

Variation of the action \eqref{2_10} with respect to $G_{AB}$ yields the gravitational equations
\begin{equation}
	R_{A}^B - \frac{1}{2}\delta_{A}^B R = \frac{1}{M^{m+2}}T_{A}^B,
\label{2_20}
\end{equation}
where the capital Latin indices run over
$A, B,... = 0, 1, 2, 3, \ldots, D$ and
\begin{equation}
	T_{A}^B = - \left[
		\left( \frac{\partial f}{\partial R}\right) R_{A}^B -
		\frac{1}{2}\delta_{A}^{B} f +
			\left( \delta_{A}^{B} g^{LM}-\delta_{A}^{L} g^{BM}
		\right)
		\left( \frac{\partial f}{\partial R}\right)_{;L;M}
	\right]
\label{2_30}
\end{equation}
is the effective energy-momentum tensor containing the nontrivial dependence on the curvature (the semicolon denotes the covariant derivative). It is seen from Eq.~\eqref{2_20}
that these equations have the structure of the Einstein equations, for which the energy-momentum conservation law is satisfied as well.

In order to describe the system under consideration, let us use the following metric:
\begin{equation}
	ds^2 = a^2(z) \eta_{\alpha\beta} dx^\alpha dx^\beta -
	d z^2-b(z) \gamma_{i j} dx^i dx^j.
\label{metr_gen}
\end{equation}
Here the Greek indices $\alpha, \beta$ refer to a four-dimensional spacetime of the brane;
$\eta_{\alpha \beta} = \{1,-1,-1,-1\}$  is the four-dimensional Minkowski metric;
$-\infty <z<\infty$ is the coordinate of the extra fifth dimension; $\gamma_{i j}$ is the $k=(m-1)$-dimensional Euclidean metric,
where the Latin indices $i, j$ refer to extra coordinates describing dimensions larger than~5; $a(z)$ and $b(z)$ are the warp factor functions depending only on $z$.
{\it Geometrically such a system can be regarded as containing two branes orthogonal to each other:
a four-dimensional Lorentzian brane and $k$-dimensional Euclidean one.}

Substituting the metric \eqref{metr_gen} into Eqs.~\eqref{2_20} and \eqref{2_30}, one can obtain (hereafter we employ units such that $M=1$)
\begin{eqnarray}
	&&	- 3 \frac{a^{\prime \prime}}{a} -
	3 \left(\frac{a^\prime}{a}\right)^2 -
	k\left[
		\frac{1}{2}\frac{b^{\prime\prime}}{b} +
		\frac{k-3}{8} \left(
			\frac{b^\prime}{b}
		\right)^2 +
		\frac{3}{2}\frac{a^\prime}{a} \frac{b^\prime}{ b}
	\right]
\nonumber\\
	&& =
		\frac{f}{2}
		-f_R \left[
			 \frac{a^{\prime \prime}}{a} +
			3 \left(\frac{a^\prime}{a}\right)^2 +
			\frac{k}{2}\frac{a^\prime}{a} \frac{b^\prime}{ b}
		\right] +
		f_{RR} R^\prime \left(
			3 \frac{a^\prime}{a} +
			\frac{k}{2}\frac{b^\prime}{ b}
		\right) + f_{RR} R^{\prime \prime} +
		f_{RRR} {R^\prime}^2	 ,
\label{eq_alpha_alpha}\\
	&&	-6 \left(\frac{a^\prime}{a}\right)^2 -
	k \left[
		2\frac{a^\prime}{a} \frac{b^\prime}{ b} +
		\frac{k-1}{8} \left(\frac{b^\prime}{b}\right)^2
	\right]
	 =  \frac{f}{2} -
		f_R \left[
			 4 \frac{a^{\prime \prime}}{a} +
			\frac{k}{2}\frac{b^{\prime \prime}}{ b} -
			\frac{k}{4}\left(\frac{b^\prime}{b}\right)^2
		\right] +
			f_{RR} R^\prime \left(
			4 \frac{a^\prime}{a} +
			\frac{k}{2}\frac{b^\prime}{ b}
		\right) 	,
\label{eq_rho_rho}\\
	&&	- 4 \frac{a^{\prime \prime}}{a}
		- 6 \left(\frac{a^\prime}{a}\right)^2
	- (k-1) \left(
		\frac{1}{2}\frac{b^{\prime \prime}}{ b} +
		2\frac{a^\prime}{a} \frac{b^\prime}{ b}
	\right) -
	\left[
		\frac{1}{2} + \frac{k(k-5)}{8}
	\right]
	\left(\frac{b^\prime}{b}\right)^2
\nonumber\\
	 && =
		\frac{f}{2}-
		f_R \left[
			 \frac{1}{2}\frac{b^{\prime \prime}}{ b} +
			\frac{k-2}{4}\left(\frac{b^\prime}{b}\right)^2 +
			2 \frac{a^\prime}{a} \frac{b^\prime}{ b}
		\right] + f_{RR} \left[
		R^{\prime \prime} +
			\left(
				4 \frac{a^\prime}{a}+\frac{k-1}{2}\frac{b^\prime}{b}
			\right) R^\prime
		\right] +
		f_{RRR} {R^\prime}^2 .
\label{eq_theta_theta}
\end{eqnarray}
Here the prime and the index $R$ by $f$ denote differentiation with respect to $z$
and $R$, respectively. Eqs.~\eqref{eq_alpha_alpha}-\eqref{eq_theta_theta} are the $\left(^\alpha_\alpha\right), \left(^z_z\right)$, and $\left(^i_i\right)$
components of Eq.~\eqref{2_20}, respectively. In turn, the scalar curvature is
\begin{equation}
	R = 8 \frac{a^{\prime \prime}}{a} +
	12 \left(\frac{a^\prime}{a}\right)^2 + k\left[
		\frac{b^{\prime \prime}}{b}	+
		\frac{k-3}{4}\left(\frac{b^\prime}{ b}\right)^2 +
		4 \frac{a^\prime}{a} \frac{b^\prime}{b}
\right].
\label{scalr_cur_gen}
\end{equation}
Thus we have the general set of ordinary differential equations \eqref{eq_alpha_alpha}-\eqref{eq_theta_theta}
and the expression \eqref{scalr_cur_gen} for the scalar curvature
written for a $D$-dimensional spacetime with arbitrary function $f(R)$.
Since these equations do not depend on the coordinate $z$ explicitly, it is always possible to shift the position of the  Lorentzian brane to the point
$z = 0$ by the corresponding transformation of coordinates.

\subsection{Six-dimensional thick brane solutions}
\label{6D_tb}

As a specific example of thick brane solutions,
in this section we focus on a special choice of $f(R)$ in the form
\begin{equation}
	f(R) =  -\alpha R^n,
\label{2_40}
\end{equation}
where the free parameters $\alpha, n > 0$. Such a function is often used in the bulk of the literature in various aspects,
including different cosmological and astrophysical applications~\cite{Nojiri:2010wj,Nojiri:2017ncd}.

Since it appears impossible to find analytic solutions of Eqs.~\eqref{eq_alpha_alpha}-\eqref{eq_theta_theta},
we perform here their numerical integration. To do this, it is necessary to choose a concrete dimension of spacetime.
In this section we consider a six-dimensional problem (i.e., the case of $k=1$), for which the metric \eqref{metr_gen} takes the form
\begin{equation}
	ds^2 = a^2(z) \eta_{\alpha \beta} dx^\alpha dx^\beta
	- dz^2 - b(z) dy^2 ,
\label{2_50}
\end{equation}
where the  coordinates $-\infty<z,y<+\infty$, and  Eqs.~\eqref{eq_alpha_alpha}-\eqref{eq_theta_theta} yield
\begin{eqnarray}
	- 3 \frac{a^{\prime \prime}}{a} -
	 \frac{1}{2}\frac{b^{\prime \prime}}{ b} +
	 \frac{1}{4}\frac{{b^\prime}^2}{ b^2} -
	\frac{3}{2} \frac{a^\prime b^\prime}{ a b} -
	3 \frac{{a^\prime}^2}{a^2} &=&
		\frac{f}{2}
		-f_R \left(
			 \frac{a^{\prime \prime}}{a} +
			3 \frac{{a^\prime}^2}{a^2} +
			\frac{1}{2}\frac{a^\prime b^\prime}{ a b}
		\right)
\nonumber\\
	&+&
	f_{RR} R^\prime \left(
		3 \frac{a^\prime}{a} +
		\frac{1}{2}\frac{b^\prime}{ b}
	\right) + f_{RR} R^{\prime \prime} +
	f_{RRR} {R^\prime}^2	 ,
\label{2_60}\\
	- 6 \frac{{a^\prime}^2}{a^2} -
	2 \frac{a^\prime b^\prime}{ a b} &=& \frac{f}{2}-
		f_R \left(
			 4 \frac{a^{\prime \prime}}{a} +
			\frac{1}{2}\frac{b^{\prime \prime}}{ b} -
			\frac{1}{4}\frac{{b^\prime}^2}{ b^2}
		\right) +
		f_{RR} R^\prime \left(
			4 \frac{a^\prime}{a} +
			\frac{1}{2}\frac{b^\prime}{ b}
		\right) 	,
\label{2_70}\\
	- 4 \frac{a^{\prime \prime}}{a} -
	6 \frac{{a^\prime}^2}{a^2} &=&
		\frac{f}{2}-
		f_R \left(
			 \frac{1}{2}\frac{b^{\prime \prime}}{ b} -
			\frac{1}{4}\frac{{b^\prime}^2}{ b^2} +
			2 \frac{a^\prime b^\prime}{ a b}
		\right) + f_{RR} \left(
			4 \frac{a^\prime}{a} R^\prime + R^{\prime \prime}
		\right) +
		f_{RRR} {R^\prime}^2 .
\label{2_80}
\end{eqnarray}
In turn, Eq.~\eqref{scalr_cur_gen} gives
\begin{equation}
	R = 8 \frac{a^{\prime \prime}}{a} + \frac{b^{\prime \prime}}{b} +
	12 \left(\frac{a^\prime}{a}\right)^2 - \frac{1}{2}\left(\frac{b^\prime}{ b}\right)^2 +
	4 \frac{a^\prime}{a} \frac{b^\prime}{b} .
\label{2_130}
\end{equation}
The form of the latter equation assumes that one can seek a solution to Eqs.~\eqref{2_60}-\eqref{2_80}
in the vicinity of the Lorentzian brane $z=0$ in the form of the series
\begin{equation}
	R(x) = R_0 + R_1 z^{\delta -2} + \cdots , \quad
a(x) = a_0 + a_1 z^\delta + \cdots , \quad
	b(x) = b_0 + b_1 z^\delta + \cdots ,
\label{2_100}
\end{equation}
where values of the expansion coefficients and of the parameter $\delta$ appearing here
must be determined as a consequence of the requirement of obtaining solutions regular on the brane.
Substitution of the expansions \eqref{2_100} in Eq.~\eqref{2_130}  gives the following expressions for the expansion coefficients:
\begin{equation}
	R_0 = 0, \quad
	R_1 = \delta (\delta - 1) \left(
		8 \frac{a_1}{a_0} + \frac{b_1}{b_0}
	\right) .
\label{2_150}
\end{equation}

\begin{figure}[h!]
\centering
  \includegraphics[width=1\linewidth]{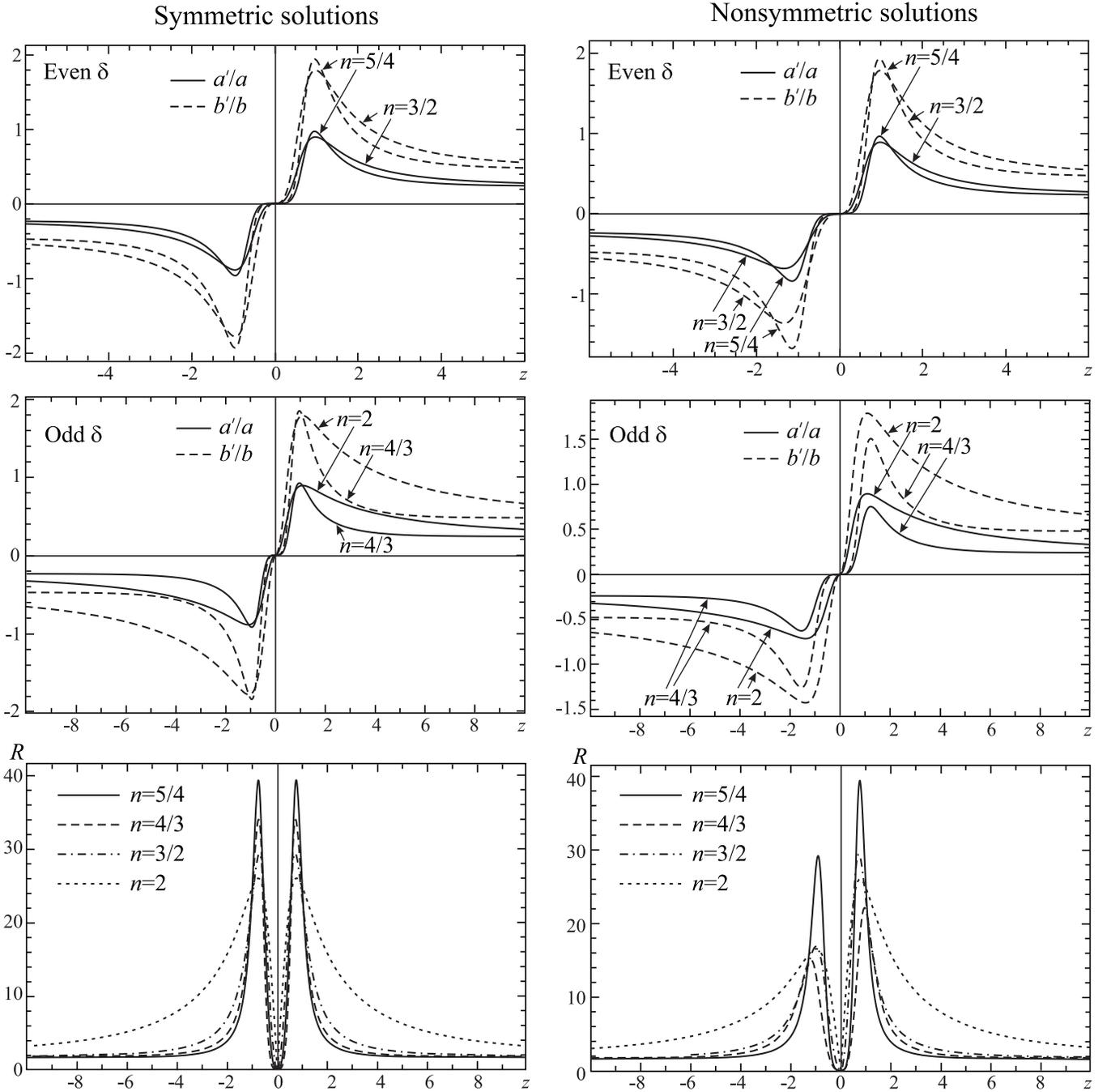}
\caption{$Z_2$-symmetric (the left panel) and nonsymmetric (the right panel) solutions passing
(in the case of $n=5/4$ and $n=4/3$) and not passing (in the case of $n=3/2$ and $n=2$) through a fixed point.
Boundary conditions are chosen as follows: {\it For the symmetric solutions,} for even $\delta$, on both sides of the brane
	$a_0 = b_0 =a_1=b_1= 1$; for odd $\delta$,  $a_{0 (+)} = b_{0 (+)} =a_{1 (+)}=b_{1 (+)}= 1$  and  $a_{0 (-)} = b_{0 (-)} = 1$, $a_{1 (-)}=b_{1 (-)}=-1$.
	{\it For the nonsymmetric solutions,}  for even $\delta$,  $a_{0 (+)} = b_{0 (+)} =a_{1 (+)}=b_{1 (+)}= 1$  and  $a_{0 (-)} = b_{0 (-)} = 1$, $a_{1 (-)}=b_{1 (-)}=0.3$;
	for odd $\delta$, when $n=4/3$,  $a_{0 (+)} = b_{0 (+)} = 1$ and $a_{1 (+)}=b_{1 (+)}=0.3$,  $a_{0 (-)} = b_{0 (-)} = 1$ and $a_{1 (-)}=b_{1 (-)}=-0.1$,
	and when $n=2$,  $a_{0 (+)} = b_{0 (+)} = 1$ and $a_{1 (+)}=b_{1 (+)}=1$,  $a_{0 (-)} = b_{0 (-)} = 1$ and $a_{1 (-)}=b_{1 (-)}=-0.5$.
For all solutions the parameter $\alpha$ from \eqref{2_40} is taken to be 1.
The indices $(+)$ and $(-)$ correspond to the right and left sides of the brane,	respectively.
}
\label{fig_ab_R}
\end{figure}

\begin{figure}[h!]
\begin{center}
	\includegraphics[width=.66\linewidth]{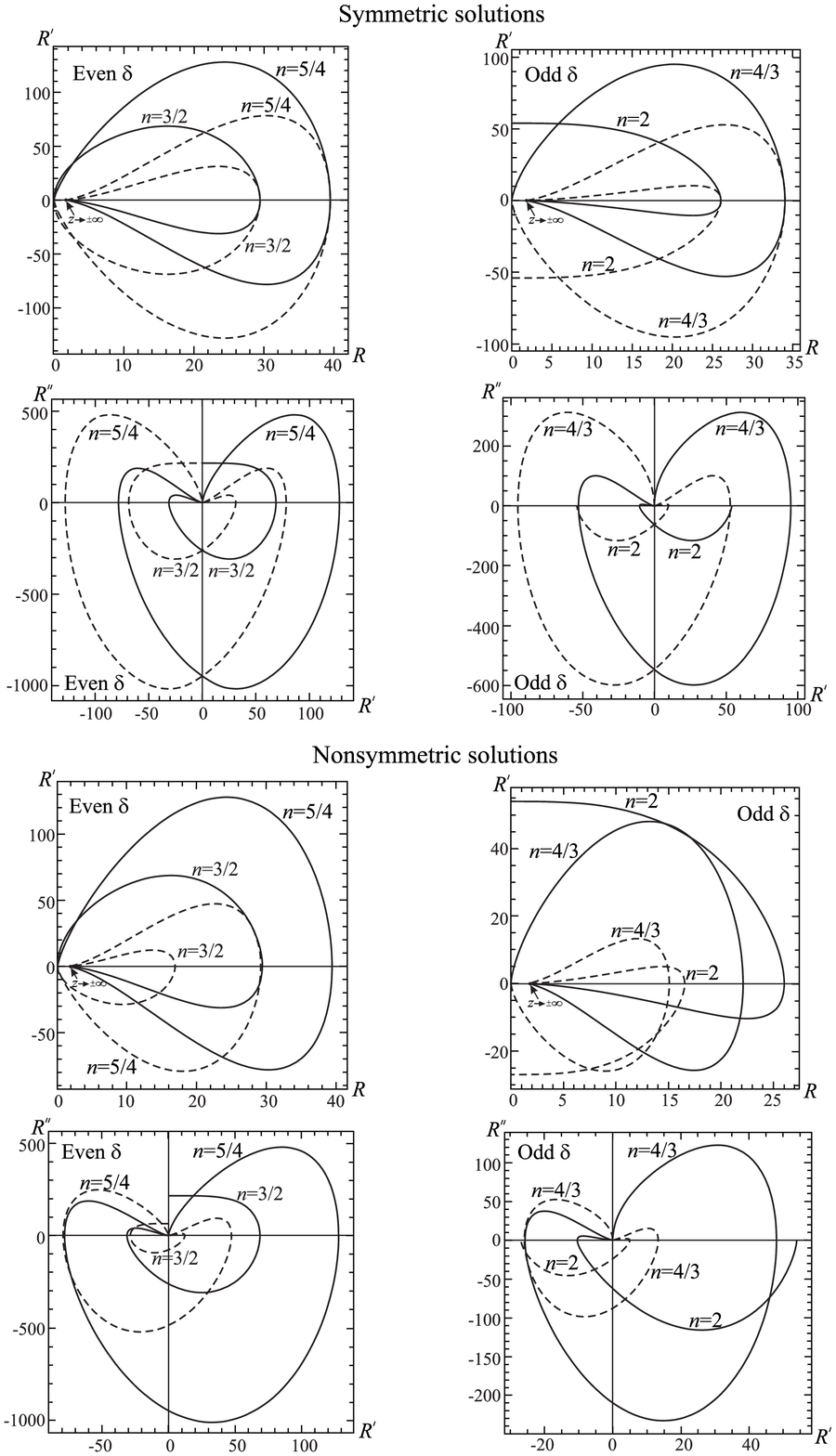}
\end{center}
\caption{The phase portraits of the solutions for the scalar curvature shown in Fig.~\ref{fig_ab_R} passing (in the case of $n=4/3$ and $n=5/4$) and not passing (in the case of $n=2$ and $n=3/2$)
through a fixed point.
The solid curves correspond to the solutions for $z>0$, the dashed lines~-- for $z<0$.
}
\label{fig_R_phasa}
\end{figure}

Next, using the expansions \eqref{2_100} in Eqs.~\eqref{2_60}-\eqref{2_80}, one can analyze the behavior of the solutions in the vicinity of the brane depending
on the values of the free parameters entering the system. Namely, it turns out
that for positive  $n$ and $\delta>2$ (this follows from the requirement of ensuring the regularity of the scalar curvature on the brane)
the leading term is the term proportional to $z^{n(\delta-2)-\delta}$. Since in general the power in the
exponent of this term can become negative for some values of  $n$ and $\delta$,
to eliminate the singularity on the brane, the coefficient in front of $z^{n(\delta-2)-\delta}$ must be equal to zero. This gives the following relation between the
parameters of the system:
\begin{equation}
	\delta = \frac{2n - 1}{n - 1},
\label{delta_n_rel}
\end{equation}
which is a necessary, albeit not sufficient, condition for the solutions on the brane to be regular. Taking this relation into account, there is the following behavior of the solutions
[we assume that the free parameters of the system are chosen so that $R_1\neq 0$, see Eq.~\eqref{2_150}]:
\begin{itemize}
  \item In the range $0<n<1$, the relation \eqref{delta_n_rel} gives $\delta<1$;
  this corresponds to singular solutions on the brane [see Eq.~\eqref{2_100}].
\item  In the range $1<n<3$,  the relation \eqref{delta_n_rel} yields $\delta> 5/2$,
thus ensuring the obtaining of solutions regular on the brane. In this case, there are two possible types of solutions:
	(i) The solutions pass through a fixed point where all functions and their derivatives go to zero on the brane.
In this case the values of the parameter $n$ lie in the range  $1<n<3/2$, corresponding to $\delta>4$.
	(ii)  The solutions do not pass through a fixed point when $3/2\leq n< 3$, corresponding to $5/2<\delta \leq 4$.
In this case, on the brane, there is either a divergence of only $R^{\prime\prime}(0)$
or a simultaneous  divergence of $R^{\prime\prime}(0)$ and $R^{\prime}(0)$. [The exceptions are two cases:
	(a) $\delta=3$ (or $n=2$) when  $R^{\prime\prime}(0)=0, R^{\prime}(0)=\text{const.}$;
	(b) $\delta=4$ (or $n=3/2$) when $R^{\prime\prime}(0)=\text{const.}, R^{\prime}(0)=0$.]
Fig.~\ref{fig_ab_R}  shows examples of the solutions for the metric functions and the scalar curvature for $n=5/4, 4/3, 3/2, 2$;
Fig.~\ref{fig_R_phasa} depicts phase portraits of the scalar curvature for the above values of $n$.	
\item
	For $n\geqslant 3$ the solutions can be regular on the brane but they diverge at some finite $z\neq 0$.
\end{itemize}

Note that the arbitrary constants $a_1$ and $b_1$ appearing in \eqref{2_100} determine the behavior of the solutions near the brane or, in fact,
specify the boundary conditions for Eqs.~\eqref{2_60}-\eqref{2_80} in the neighborhood of the Lorentzian brane.
Depending on the values of these constants and of the parameter  $\delta$, the solutions
can be either symmetric relative to the axis $z=0$ ($Z_2$-symmetric) or nonsymmetric. This can be seen from the behavior of the solution
in the neighborhood of the brane \eqref{2_100} where the values of both the functions $a, b$ and their derivatives will depend, generally speaking,
on that where we are situated~-- to the right ($z>0$) or the left ($z<0$) side of  the brane.

Since we seek regular solutions, it is necessary to make a corresponding choice of $\delta$ for given $n$, defined in Eq.~\eqref{delta_n_rel}.
$Z_2$-symmetric solutions are then possible in the following cases:
\begin{itemize}
  \item
	If  $\delta$  is an even number then $a_1$ and $b_1$ must be positive and equal
	on either side of the brane, i.e., $a_{1(+)}=a_{1(-)}$ and $b_{1(+)}=b_{1(-)}$.
 \item If  $\delta$  is an odd number then $a_1$ and $b_1$ must be equal in modulus and
 have different signs on the left and on the right sides of the brane (plus at $z>0$ and minus at $z<0$), i.e., $a_{1(+)}=-a_{1(-)}$ and $b_{1(+)}=-b_{1(-)}$.
 Examples of the corresponding regular solutions are shown in the left panel of Fig.~\ref{fig_ab_R}. We emphasise that in this case
 the spaces on both sides of the brane are described by different metrics.
 \end{itemize}

If the above conditions are not fulfilled, regular solutions either are absent or they are not $Z_2$-symmetric.
Examples of the latter solutions are given in the right panel of Fig.~\ref{fig_ab_R}.
In this case the spaces on either side of the brane are also described by different metrics.
But, as opposed to a thin brane, where a derivative discontinuity of the metric  function occurs,
in the case of the thick brane considered here, the metric functions and their derivatives always remain smooth functions.

Finally, the asymptotic form of the solutions in the range $1<n<3$ is as follows:
\begin{equation}
	a\approx a_\infty e^{l_a |z|}, \quad b\approx b_\infty e^{l_b |z|},
\label{asymp}
\end{equation}
where $a_\infty, b_\infty, l_a, l_b$ are some parameters whose numeric values are determined by the free parameters of the system under consideration
[see Eq.~\eqref{lb_asymp_k} below].
The corresponding asymptotic value of the scalar curvature  \eqref{2_130} is
\begin{equation}
	R_{(k=1)} \approx \frac{1}{2} \left(
		40 l_a^2 + 8 l_a l_b + l_b^2
	\right).
\label{R_asymp_6}
\end{equation}
For the solutions considered here, this quantity is positive, i.e., the spacetime is asymptotically (as $z\to \pm\infty$) anti-de Sitter one.

\subsection{Gibbons-Hawking-York boundary term }

Let us now calculate the Gibbons-Hawking-York boundary term $S_{\text{GHY}}$ from
\eqref{GHY_term} for the above-considered six-dimensional case with the metric \eqref{2_50}.
For higher-dimensional modified gravity the corresponding boundary term may drastically change the brane dynamics,
as it was demonstrated in Refs.~\cite{Cvetic:2001bk,Nojiri:2004bx}.

As a boundary, we take two hypersurfaces with constant $t=t_1, t_2$ and a hypersurface at $z\equiv z_0=\text{const}$.
In this case the induced metric is
$$
h_{ab}=\{a^2(z_0),-a^2(z_0),-a^2(z_0),-a^2(z_0),-b(z_0)\}
$$
(the indices $a,b=0,1,2,3,6$) and correspondingly  $\sqrt{|h|}=a^4(z_0)\sqrt{b(z_0)}$. A unit normal is $n_A=\partial_A z$. Then we have
[here we employ the full metric \eqref{2_50}]
$$
K=n^A_{;A}=\frac{\partial}{\partial x^A}\left(g^{AB}n_B\right)+\Gamma^A_{CA}g^{CB}n_B=-\left[4\frac{a^\prime(z_0)}{a(z_0)}+\frac{1}{2}\frac{b^\prime(z_0)}{b(z_0)}\right].
$$
As a result, for the integral \eqref{GHY_term}, we have
\begin{equation}
\begin{split}
	\oint \sqrt{|h|}F_R K dx^\alpha dy = & - \int_{t_1}^{t_2} dt
	\int_{-\left( x^1 \right)_1}^{\left( x^1 \right)_2} d x^1
	\int_{-\left( x^2 \right)_1}^{\left( x^1 \right)_2} d x^2
	\int_{-\left( x^3 \right)_1}^{\left( x^3 \right)_2} d x^3
	\int_{-z_1}^{z_2} dz
\\
	&
	\times\left[
		4\frac{a^\prime(z_0)}{a(z_0)}+\frac{1}{2}\frac{b^\prime(z_0)}{b(z_0)}
	\right] F_R(z_0) a^4(z_0)\sqrt{b(z_0)}
\\
	&
	=\left( t_2 - t_1 \right)
	\left[ \left( x^1 \right)_2 + \left( x^1 \right)_1 \right]
	\left[ \left( x^2 \right)_2 + \left( x^2 \right)_1 \right]
	\left[ \left( x^3 \right)_2 + \left( x^3 \right)_1 \right]
	\left( z_2 + z_1 \right)
\\
	&
	\times
\left[
		4\frac{a^\prime(z_0)}{a(z_0)}+\frac{1}{2}\frac{b^\prime(z_0)}{b(z_0)}
	\right]	
F_R(z_0) a^4(z_0)\sqrt{b(z_0)}.
\end{split}
\label{GHY_calc}
\end{equation}
This surface integral should be added to the action to ensure the regularity under its variation.

Using the asymptotic solutions \eqref{asymp}, we have $a^\prime/a\to \pm l_a, b^\prime/b\to \pm l_b$ (the plus sign  is taken for $z_0>0$, the minus sign~-- for $z_0<0$),
where the constants $l_a$ and $l_b$ to the right and to the left of $z_0=0$ are not equal to each other for the nonsymmetric solutions.
Correspondingly, the term in square brackets in Eq.~\eqref{GHY_calc} always gives a finite contribution to the integral.

\subsection{Higher-dimensional cases}

For the systems with $k>1$, numerical computations indicate that the qualitative behavior of the solutions remains essentially the same as that in the case of
 codimension 1 configurations considered in Sec.~\ref{6D_tb}. Here, regular solutions do exist in the range  $1<n<D/2$,
 and the boundary conditions \eqref{2_100}, the relation \eqref{delta_n_rel}, and the asymptotic behavior of the solution  \eqref{asymp} remain valid.
 As a result, asymptotically, one has an anti-de Sitter spacetime with the scalar curvature
 \begin{equation}
	R_{(k)} \approx \frac{1}{2} \left[
		40 l_a^2 + 8 k l_a l_b + \frac{k(k+1)}{2} l_b^2
	\right].
\label{R_asymp_k}
\end{equation}
The parameters  $l_a$ and $l_b$ entering here are defined as follows:
\begin{equation}
l_b^{2(n-1)}=\frac{4^{n-1}}{\alpha}\frac{k^3+12k^2+47k+60}{(k-2n+5)\left(k^2+9k+20\right)^n} \quad \text{and} \quad l_a=\frac{l_b}{2}.
\label{lb_asymp_k}
\end{equation}
Notice that in the range  $1<n<D/2$ the right-hand side of the expression for $l_b$ remains always positive.

\section{Limiting solutions for $k\gg 1$}

In this section we consider a situation where the Euclidean brane, and correspondingly all of spacetime,  has a very large dimension,
i.e., the limiting case of $k\gg 1$. In this case, numerical computations indicate that one can distinguish two regions in the bulk.
In the former, within the framework of modified gravity \eqref{2_40}, the general equations
\eqref{eq_alpha_alpha}-\eqref{eq_theta_theta}  reduce to the following two approximate equations
(the terms appearing here give leading contributions):
 \begin{eqnarray}
\label{gen_eqs_approx_1}
	&&f_{RR}R^{\prime\prime}+\frac{k}{2}f_{RR}R^\prime\frac{b^\prime}{b}+\frac{f}{2}=0,\\
&&\frac{k}{2}f_R\frac{b^{\prime\prime}}{b}-\frac{k}{2}f_{RR}R^\prime\frac{b^\prime}{b}-\frac{f}{2}=0.
\label{gen_eqs_approx_2}
\end{eqnarray}
Taking into account \eqref{2_40}, they yield
\begin{eqnarray}
\label{part_eqs_approx_1}
	&&R^{\prime\prime}+\frac{k}{2}R^\prime\frac{b^\prime}{b}+\frac{R^2}{2n(n-1)}=0,\\
&&\frac{b^{\prime\prime}}{b}-(n-1)\frac{R^\prime}{R}\frac{b^\prime}{b}-\frac{R}{k \,n}=0.
\label{part_eqs_approx_2}
\end{eqnarray}
In turn, in the limit $k\gg 1$, the scalar curvature \eqref{scalr_cur_gen} takes the form
\begin{equation}
	R \approx k\left[\frac{b^{\prime\prime}}{b}+\frac{k}{4}\left(\frac{b^{\prime}}{b}\right)^2\right].
\label{R_asymp_approx}
\end{equation}
Substituting this expression into \eqref{part_eqs_approx_2}, for the physically interesting case of $n\ll k$,
one can obtain the equation
 \begin{equation}
\frac{y^{\prime\prime}}{y}-n\left(\frac{y^\prime}{y}\right)^2=C_1 \left(y^\prime\right)^{1/n},
\label{y_asymp_approx}
\end{equation}
where $y=b^{k/[4(1-n)]}$ and $C_1$ is an integration constant. For the values of  $n$ used in the paper this equation can be integrated
analytically in terms of  special functions (we do not show these solutions here to avoid overburdening the text).

\begin{figure}[t]
\centering
  \includegraphics[height=5.8cm]{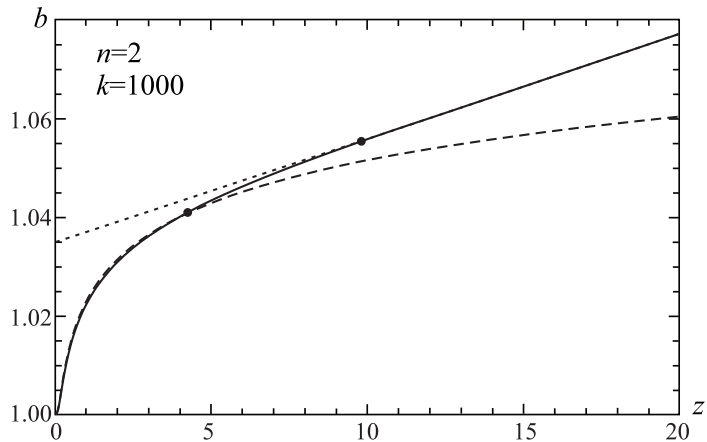}
\caption{A comparison of the exact numerical solution of Eqs.~\eqref{eq_alpha_alpha}-\eqref{eq_theta_theta}
(solid line) with the approximate analytic solutions \eqref{b_approx_n_2}
(long-dashed line) and \eqref{b_approx_k} (short-dashed line) for the case of $n=2$ and $k=1000$ with $\alpha=b_0=1$.
The transitional region lies between the two  bold dots.
}
\label{fig_b_large_k}
\end{figure}

The numerical analysis shows that for the values of the free parameters of the system used here Eq.~\eqref{y_asymp_approx}
is valid only in a relatively small region near the Lorentzian brane. When $z$ increases further,
Eqs.~\eqref{eq_alpha_alpha}-\eqref{eq_theta_theta} contain already only two leading terms, yielding the following approximate equation:
\begin{equation}
\frac{k^2}{8}\left(\frac{b^\prime}{b}\right)^2+\frac{f}{2}=0,
\label{asymp_approx_2}
\end{equation}
and Eq.~\eqref{scalr_cur_gen} reduces to
$
R\approx \left(k^2/4\right)\left(b^\prime/b\right)^2.
$
Using the latter expression in \eqref{asymp_approx_2} and taking into account \eqref{2_40},
one can obtain the following solution of Eq.~\eqref{asymp_approx_2}:
\begin{equation}
b=b_\infty e^{l_{b} |z|} \quad \text{with} \quad l_{b}=\frac{2}{\alpha^{1/[2(n-1)]}}\frac{1}{k}.
\label{b_approx_k}
\end{equation}
This solution corresponds to the asymptotic solution \eqref{asymp},
and the parameter $l_b$ is obtained from \eqref{lb_asymp_k} in the large-$k$ limit. As a result,
the exact solution is approximated by a solution of Eq.~\eqref{y_asymp_approx} in regions close to the Lorentzian brane
and by the expression \eqref{b_approx_k} in large-$z$ regions.
Also, there is a transitional region, for the description of which  it is necessary to take into account all the terms
entering the exact equations \eqref{eq_alpha_alpha}-\eqref{eq_theta_theta} (cf. Fig.~\ref{fig_b_large_k}).

As an example, Fig.~\ref{fig_b_large_k} shows a comparison of the exact numerical solution with the approximate analytic solutions
for the case of $n=2$. In this case Eq.~\eqref{y_asymp_approx} can be integrated to yield
\begin{equation}
b=b_0\left(1+C_2 z^3\right)^{4/k},
\label{b_approx_n_2}
\end{equation}
where $C_2$ is an integration constant. This solution adequately approximates the exact solution up to $z\sim 5$.
For larger values of $z$, there is a transitional region for $5\lesssim z \lesssim 10$, and for $z \gtrsim 10$
the exact solution is well approximated by the asymptotic expression \eqref{b_approx_k}.
One might expect that a qualitatively similar behavior will occur for other values of the parameters
 $n$ and $\alpha$ as well.

Notice also that in the large-$k$ limit the asymptotic scalar curvature \eqref{R_asymp_k} is described by the expression
$$
R_{(k\gg 1)}\approx \alpha^{1/(1-n)}
$$
which does not already involve $k$. Thus the properties of the asymptotic anti-de Sitter space are determined only by the parameters of
modified gravity \eqref{2_40}.

\section{Trapping of scalar fields}
In constructing brane models, it is assumed that zero modes of various matter fields corresponding to particles or fields in the Standard Model should be confined on the brane.
Here we demonstrate the possibility of trapping of scalar fields on the Lorentzian brane under consideration.
Following Ref.~\cite{Dzhunushaliev:2009dt}, let us take the test complex scalar field
$\chi$ with the mass $m_0$, whose Lagrangian is
$$
	L_{\chi} = \frac{1}{2} \partial_A \chi^{*}\partial^A\chi -
	\frac{1}{2}m_0^2 \chi^*\chi.
$$
Varying this Lagrangian with respect to $\chi^*$, we have the following equation for the scalar field:
\begin{equation}
\label{tf_eq}
	\frac{1}{\sqrt{^{(D)} G}}
	\frac{\partial}{\partial x^A}
		\left(\sqrt{^{(D)} G} \,G^{AB} \frac{\partial \chi}{\partial x^B}
	\right) = - m_0^2 \chi,
\end{equation}
where $\chi$ is a function of all coordinates, $\chi=\chi(x^A)$, and $\sqrt{^{(D)} G}=a^4 b^{k/2}$. We will seek a solution to this equation in the form
$$
  \chi (x^{A}) = X(z) \exp (-ip_{\mu }x^{\mu }),
$$
where $p_\mu=(E, \vec{p}\,)$ are the canonically conjugate momenta. Substituting this expression into Eq.~\eqref{tf_eq}, we have
$$
  X^{\prime\prime} + \left(
  	4 \frac{a^{\prime}}{a} + \frac{k}{2} \frac{b^{\prime}}{b}
  \right)X^{\prime} + (p^{\mu }p_{\mu } - m_{0}^2 ) X = 0,
$$
where  $p^{\mu}p_{\mu}=a^{-2}\left(E^2-\vec{p}\,^2\right)$ and the prime denotes differentiation with respect to $z$.
Next, using the asymptotic solutions~\eqref{asymp} and neglecting  the term  $p^{\mu}p_{\mu}$ in comparison with $m_0^2$, we obtain
$$
	X^{\prime\prime} + \left(4 l_a +
	\frac{k}{2}l_b\right) X^{\prime} - m_0^2  X = 0,
$$
whose asymptotically decaying solution is
\begin{equation}
\label{asymp_sol}
	X_{\infty} \approx C \exp \left[
		-\frac{8 l_a + k l_b +
		\sqrt{\left(8 l_a + k l_b\right)^2 + 16 m_0^2}}{4} |z|
\right],
\end{equation}
where $C$ is an integration constant.

In order to ensure the trapping of matter on the brane, we must require finiteness of the field energy
per unit 3-volume of the brane, i.e.,
$$
	E_{\rm tot}[\chi] = \int \limits_{0}^{\infty} T^0_0 \sqrt{^{(D)}G} \,d z
  = \int\limits_{0}^{\infty}
  a^4 b^{k/2}\left[
  	a^{-2}(E^2 + \vec{p}\,^2)X^2 + m_0^2 X^2 + X^{\prime 2}
	\right] d z < \infty,
$$
and the finiteness of the norm of the field $\chi$
$$
	||\chi||^2 = \int\limits_{0}^{\infty }\sqrt{^{(D)}G}\,\chi^*\chi\,d z
  = \int_{0 }^{\infty } a^{4} b^{k/2} \,X^2 \,d z.
$$
From the asymptotic solutions \eqref{asymp} and \eqref{asymp_sol}, it is evident
that both $E_{\rm tot}$ and $||\chi||$ converge. This corresponds to the fact that the test scalar field is confined on the brane and the  localized solutions obtained in Sec.~\ref{eqs} may be interpreted as thick brane solutions.

\section{Conclusion}

We have considered $D$-dimensional thick brane configurations within the framework of pure $f(R)\sim R^n$ gravity (without matter).
The systems under consideration consist of two branes orthogonal to each other:
the four-dimensional Lorentzian brane and $k$-dimensional Euclidean one. For such systems, it is shown that vacuum regular solutions do exist
only in the range  $1<n<D/2$. Depending on the values of the parameter $n$ and boundary conditions, there are
both $Z_2$-symmetric and nonsymmetric solutions that can pass or not pass through a fixed point lying on the Lorentzian brane. For all solutions considered here
the spacetime is asymptotically (far from the Lorentzian brane) anti-de Sitter.

For the case when the Euclidean brane has a very large dimension ($k\gg 1$) we have found approximate analytic solutions that adequately represent
exact numerical solutions. In future, one can calculate corrections to these solutions by expanding in powers of
$1/k$.

Also, a consideration of the behavior of a test scalar field in the bulk has shown that such a field is confined on the brane
independently of the dimension of the external space, and the trapping is purely gravitational.
Note that in a similar spirit one can get the trapping of gravitational field. However, the detail of that complicated study goes beyond  the scope of this paper.

It is interesting to note that the thick branes solutions obtained here have an attractive interpretation in the spirit of
$T^2$ wormhole solutions found in Refs.~\cite{GonzalezDiaz:1996sr,Dzhunushaliev:2019qze}. The coordinates
$x^\alpha = x^2, x^3$ in the metric
$\eta_{\alpha\beta} dx^\alpha dx^\beta$ in \eqref{metr_gen} can be factorized with the periods $x_0^2, x_0^3$.
This leads to the fact that the space spanned on the coordinates $x^2, x^3$ becomes a $T^2$ torus.
In this case the brane with the coordinates  $x^\alpha = x^0, x^1, x^2, x^3$ can be regarded as a trivial $T^2$ wormhole embedded in the bulk with the
coordinates $\{x^\alpha, z, x^i\}$. This wormhole does not asymptotically flat, and from the geometric point of view it looks like a tube with a constant cross section
in the form of a torus with radii $x_0^2, x_0^3$.

\section*{Acknowledgments}
V.D. and V.F. gratefully acknowledge support provided by Grant No.~BR05236494 in Fundamental Research in Natural Sciences
by the Ministry of Education and Science of the Republic of Kazakhstan. They are also grateful to the Research Group Linkage Programme
of the Alexander von Humboldt Foundation for the support of this research
and would like to thank the
Carl von Ossietzky University of Oldenburg for hospitality
while this work was carried out.
This work is also supported by MINECO (Spain), FIS2016-76363-P, and by project 2017 SGR247 (AGAUR, Catalonia) (S.D.O).

\appendix
\section{The surface action}
\label{sur_act}
In the present paper, we have considered a thick brane embedded in a $D$-dimensional bulk.
For this case, it is assumed that in the bulk gravity is modified. In some sense, one can consider the opposite problem:
gravity is modified only on the brane but it remains Einsteinian in the bulk. In this case the surface action  $S_1$ from \eqref{2_10} is
\begin{equation}
	S_1 = \frac{1}{2} \int d^4 x
	\sqrt{|h|}\, F\left( R_{(4)} \right),
\label{4_10}
\end{equation}
where $R_{(4)}$ is the scalar curvature on the brane. Such brane models first arose when considering models with infinite extra dimensions
(see, e.g., Refs.~\cite{arkani1,arkani2,gog,rs,fermi1,fermi11}). Our purpose here is to construct branes in vacuum, i.e., in the absence of mater;
hence, in constructing a thin brane model, it is reasonable to take the corresponding action in the form \eqref{4_10}.

It is evident that on a four-dimensional plane symmetric brane the metric is flat, and therefore the Ricci tensor is zero, and hence the scalar curvature
$R_{(4)} = 0$. Therefore, on the flat brane, the Lagrangian for the surface action $S_1$ is constant:
$$
	F\left( R_{(4)} \right) = F\left( 0 \right).
$$

By varying with respect to the metric, the surface equation (the Israel junction conditions~\cite{Israel:1966rt}) is
\begin{equation}
	K_{\mu \nu} - K h_{\mu \nu} =  t_{\mu \nu},
\label{4_20}
\end{equation}
where the extrinsic curvature $K_{\mu \nu}$ is defined as
$K_{\mu \nu} = \nabla_\mu n_\nu$, $K = h^{\mu \nu} K_{\mu \nu}$, and the surface energy-momentum tensor is
\begin{equation}
	t_{\mu \nu} = \frac{2}{\sqrt{-h}} \frac{\delta S_1}{\delta h^{\mu \nu}}.
\label{4_30}
\end{equation}
By varying of the surface action \eqref{4_10} with respect to the four-dimensional metric $h_{\mu \nu}$, we have
\begin{equation}
	t_{\mu \nu} = R_{\mu \nu} - \frac{1}{2} h_{\mu \nu} R -
	 T_{\mu \nu} =
			a^2(0) \eta_{\mu \nu} \left[
			\frac{f}{2} - \left( f_R \right)^{, \alpha}_{, \alpha}
		\right] + \left( f_R \right)_{, \mu , \nu}.
	\label{4_40}
\end{equation}
Here the expression for $T_{\mu \nu}$ is given by Eq.~\eqref{2_30} (for a four-dimensional case), and we have taken into account that the Ricci tensor
and the scalar curvature are equal to zero for a four-dimensional metric on the brane.

Taking into account the junction conditions \eqref{4_20}, one can construct a thin brane model where gravity is modified only on the brane but not in the bulk.
It would be interesting to compare the results of calculations for this case with those obtained in the present paper for the case when gravity is modified in the bulk.
But this is a separate problem which we plan to address in future work.

\end{document}